\def \cm{~\rm{cm}}
\def \s{~\rm{s}}
\def \km{~\rm{km}}

\def \K{~\rm{K}}
\def \g{~\rm{g}}
\def \G{~\rm{G}}
\def \AU{~\rm{AU}}
\def \erg{~\rm{erg}}

\def \yr{~\rm{yr}}

\documentclass[12pt,preprint]{aastex}
\usepackage{natbib}
\usepackage{epsfig}
\begin{document}   % Leave intact

\title{Why Magnetic Fields Cannot be the Main Agent Shaping Planetary Nebulae}
%%p \titlemark{Shaping Planetary Nebulae and Related Objects}

\author{Noam Soker\altaffilmark{1}}
\affil{Department of Physics,
Technion$-$Israel Institute of Technology, Haifa 32000, Israel,
Email: soker@physics.technion.ac.il}

%
% Press releases:
%  http://www.stefan-jordan.de/pr20041993.htm
%-----------------------------------------------------------------------
%                  Abstract
%-----------------------------------------------------------------------

\begin{abstract}
An increasing amount of literature reports the detection of
magnetic fields in asymptotic giant branch (AGB) stars and in
central stars of planetary nebulae (PNs). These detections lead to
claims that the magnetic fields are the main agent shaping the
PNs. In this paper, I examine the energy and angular momentum
carried by magnetic fields expelled from AGB stars, as well as
other physical phenomena that accompany the presence of large
scale fields, such as those claimed in the literature. I show that
a single star cannot supply the energy and angular momentum if the
magnetic fields have the large coherent structure required to
shape the circumstellar wind. Therefore, the structure of
non-spherical planetary nebulae cannot be attributed to
dynamically important large scale magnetic fields. I conclude that
the observed magnetic fields around evolved stars can be
understood by locally enhanced magnetic loops which can have a
secondary role in the shaping of the PN. The primary role, I
argue, rests with the presence of a companion.
\end{abstract}


% ========================================
\section{INTRODUCTION}
% ========================================
Only a small fraction of the observed planetary nebula (PNs)
population has spherical structure.
Most PNs possess axisymmetrical structures, with rich variety
of morphologies.
One of the major open questions regarding PNs is the
mechanism responsible for their shaping
(Balick \& Frank 2002).
The key question in any model is whether a binary companion,
stellar or substellar, is required for the shaping of PNs,
or whether stellar rotation and magnetic fields can be alone
responsible for the observed morphologies.

In a series of papers (summarized in Soker 2004) I have claimed
that a binary companion, stellar or substellar, must be present
to form a non-spherical PN, with
different binary companions influencing the mass loss geometry
from PN progenitors in different ways (e.g., accreting
mass and blowing jets, spinning up the envelope or
colliding with the AGB core - sec. 2 of Soker 2004).

On the other hand, magnetic fields are still perceived in the literature
as plausible physical mechanisms which can be the {\it sole}
responsible for the shaping of PNs.
This idea has been reinforced by the discovery of magnetic fields
in AGB stars (e.g., Etoka \& Diamond 2004, Bains et al.\ 2004b)
and more recently in actual central stars of PN (Jordan et al. 2005).

In this paper I explain how these findings, interesting that they are,
remain far from providing an explanation of how PNs acquired their
axi-symmetric shapes. In particular angular momentum and energy
considerations lead to concluding that global magnetic fields,
such as those claimed in the literature, cannot be present in AGB stars
(Section 3.1). I also show that other physical considerations such
as magnetic energy (section 3.2) magnetic stress (Section 3.3),
and ambipolar diffusion (Section 3.4),
make PN shaping by magnetic fields unlikely. I finally conclude (Section 4),
arguing that {\it local} magnetic fields are consistent with those observed
and that they can take a side role in the shaping of local regions inside
the main nebula.
The burst of strongly polarized OH maser emission
in the proto-PN OH17.7-2.0 supports a local nature of the
polarized maser emission (Szymczak \& Gerard 2005).
The most likely primary shaping agent, however,
remains with the presence of companions.
The binary model is supported by recent claims for
more PNs harboring central binary systems
(De Marco et al. 2004; Hillwig 2004; Sorensen \& Pollacco  2004).

% ===============================================
\section{MAGNETIC FIELDS}
% ===============================================
% ===============================================
\subsection{The detection of magnetic fields}
% ===============================================

There are now several detections of magnetic fields in PNs and around
asymptotic giant branch (AGB) stars.
The most convincing measurements come from polarized maser emission
(e.g., Zijlstra et al.\ 1989; Kemball \& Diamond 1997; Szymczak et al.\
1998; Szymczak \& Gerard 2004, 2005; Miranda et al.\ 2001;
Vlemmings et al.\ 2002, 2005; Diamond \& Kemball 2003;
Bains et al.\ 2003, 2004a,b; Jordan et al.\ 2005).
Crudely, these works find the strength of the magnetic field to
be $\sim 1 \G$ in SiO maser clumps near the AGB stellar surface
at $r \sim 1 \AU$, decreasing to $\sim 10^{-4}-10^{-3} \G$ in
H$_2$O and OH masers clumps at $r \sim 100-1000 \AU$.
The geometry varies from large scale ordered polarization
to disordered fields, depending on the object.
Large uncertainties may be involved in determining the magnetic
field strength.
For example, Richards et al.\ (2005) find discrepancies between the
EVN/global VLBI and MERLIN data.
It may even be that in some cases the strength of the magnetic field in
maser spots is overestimated,
 i.e., non-Zeeman interpretation are possible in SiO masers
(Wiebe \& Watson 1998).
Variable X-ray luminosity, as was found in Mira A (Karovska et al.\ 2005),
can also hint at the presence of magnetic activity, although
the field strength and geometry cannot be inferred.

More recently, Jordan et al. (2005) made the claim that magnetic fields
were found in actual central star of PN, as opposed to their precursors.
They find conclusive evidence for magnetic fields in 2 central
stars of PNs, with possible detections in an additional two PNs. From
these detection they conclude that magnetic fields have shaped the PNs.

Some of these papers jump from their magnetic field detections, to the
conclusion that magnetic fields shape PNs (e.g., Etoka \& Diamond 2004;
Bains et al.\ 2004b; Jordan et al.\ 2005), but more often than not,
no justification of the mechanism is provided.

In Soker (2002), I showed that the detected magnetic fields
do not necessarily play a role in globally shaping PNs.
In that paper and in Soker \& Kastner (2003) it is
argued that the magnetic fields can only be
of small coherence lengths; such small scale fields
might result from stellar magnetic spots or flares,
or from jets blown by an accreting companion.
These are different from the global fields needed to shape the
entire PN.
In the solar wind, magnetic pressure exceeds thermal
pressure only in magnetic clouds  (e.g., Yurchyshyn et al.\ 2001),
which are formed by discrete mass loss events from the sun.
Soker \& Kastner (2003) suggest that the maser spots with strong
magnetic fields which are observed around some AGB stars might be
similar in nature to the magnetic clouds in the solar wind,
in that they represent local enhancements of the magnetic field.
 The fast variation in the polarization of the OH maser emission
in the proto-PN OH17.7-2.0 (Szymczak \& Gerard 2005) seems to support
a local explanation for a strong magnetic field.
This magnetic field topology is drawn schematically in Figure 1.
Locally strong magnetic fields, formed by local dynamo based
on the convective envelope, were suggested also to exist
on the giant star Betelgeuse (Dorch 2004; Dorch \& Freytag 2002).
In both AGB stars and Betelgeuse, the locally strong magnetic fields
are due to the strong convection in the envelope.
Therefore, the magnetic field detected in maser spots might
originate from one of the following sources, rather than
being a large scale field:
(1) Locally (but not globally) strong magnetic field spots
on evolved stars.
(2) A magnetically active main sequence companion.
(3) An accretion disk around a companion, which amplifies
the magnetic field of the accreted gas; the magnetic field can
be transported to large distances by jets.

In this paper I reinforce the argument presented by Soker (2002) by
demonstrating in a physical way how magnetic fields of the type detected
cannot play a dominate role in globally shaping PNs, although they might
still play a role in local shaping
(namely, on physical scales much shorter than the PN radius).

\subsection{The role of magnetic fields in shaping PN morphologies}
Models where magnetic fields play a role in shaping the circumstellar
matter belong to one of four categories.
\begin{enumerate}
\item Magnetic fields deflect the flow close to the stellar surface,
i.e., the magnetic pressure and/or tension are dynamically
important already on the AGB (or post-AGB) stellar surface
(e.g., Pascoli 1997; Matt et al.\ 2000; Blackman et al.\ 2001;
Garc\'{\i}a-Segura et al.\ 2005).
\item The magnetic field is weak close to the AGB stellar surface, but
it plays a dynamical role at large distances from the star
(e.g., Chevalier \& Luo 1994; Garc\'{\i}a-Segura 1997;
Garc\'{\i}a-Segura et al.\ 1999).
\item Local magnetic fields on the AGB stellar surface
are enhanced in cool spots and filaments.
The dust formation rate, and hence the mass loss rate, is enhanced above these
cool spots and filaments (section 2 in Soker 2004, and references therein).
Magnetic fields never become dynamically important on a large scale.
This mechanism can operate even for very slowly rotating
AGB envelopes, and spin-up by companions even as small as
planets is enough to account for the required rotation velocity.
(This process may lead to the formation of moderate elliptical PNs
 - those with small departure from sphericity, but can't
account for lobes, jets, etc.).
\item Magnetic fields play a dominant role in the launching of jets
from accretion disks, either around stellar companions or
around the central star. The action of the jets shapes the PN.
\end{enumerate}

Soker (1998) and Soker \& Zoabi (2002) conclude that models
(1) and (2) above, which invoke strong stellar-magnetic field
shaping cannot work without a companion constantly spinning-up the
envelope. It is the companion that has the most influential
effects on the mass loss process, not the magnetic field per se.
At most, the magnetic field plays a secondary role. Models for
magnetic activity during the post-AGB phase, have in addition the
problem of explaining the bipolar outflow from systems where the
progenitor is still an AGB star. Another problem of models based
on process (2), is to explain point symmetric PNs, which result
form precession. The previous arguments against the magnetic
models (1) and (2) above are as follows.
($i$) In some models of
type (1) above, the assumed rotation profile in the envelope is too
fast or not realistic. It is not realistic in the sense that a
large angular velocity gradient is assumed to exist inside the
envelope. But the same strong magnetic field inside the envelope
enforces a uniform rotation in a short time (Soker \& Zoabi 2002).
The rotation is too fast in the sense that the required envelope
rotational velocity is much above what a single AGB star can
possess (Soker 1998).
($ii$) There is no good explanation to the
transition from spherical to non-spherical mass loss geometry
before the star leaves the AGB (Soker \& Zoabi 2002).
($iii$) Models of type (1) do not consider the role of radiation pressure
on dust as the mechanism driving the high mass loss rate at the
end of the AGB.

In processes (3) and (4) the magnetic field plays an auxiliary role
and is not directly responsible for the shaping.
In process (3) the mass loss geometry is
determined by radiation pressure on dust.
The magnetic field originated by the cool spots, only leads to higher
concentration of spots in the equatorial plane. This process is more
similar to models based on non-radial pulsation of the envelope,
where photospheric behavior shapes the wind, rather than models based on
dynamical magnetic fields.

{{{ The jet shaping in category 4 is clearly
different from the first three types of models
\newline
($i$) The energy source of the magnetic field is different.
In the first three categories the main
energy source is the {\it nuclear} burning in the AGB (or post-AGB) core.
It drives the convective envelope that amplifies the magnetic field.
In the 4th type (category), it is {\it gravitational} energy of the accreted gas
that is partially converted to magnetic energy.
This is a fundamental difference.
\newline
($ii$) In the first three groups the magnetic field is amplified by the AGB,
or in some models by the post-AGB, star. In the 4th group the
magnetic field is amplified by an accretion disk around the companion or
around the central post-AGB star.
\newline
($iii$) In most models of the first three categories the magnetic fields directly
influence (shape) the nebular material, as it leaves the AGB star, or at large
distances.
In the 4th category, that of jet shaping, the magnetic field influences only
a small portion of the mass that ends in the nebula (if it does at all).
Only in some models of the first category the magnetic field of the post AGB
stars shapes two fast jets which contain a small amount of mass.
\newline
($iv$) As stated above, in the fourth class of models the magnetic field is
not a necessary ingredient; it is enough that two jets (or a collimated fast wind
[CFW]) are launched, whether by magnetic fields, by thermal pressure,
or by other mechanisms.
\newline
($v$) As I have shown in several papers (Harpaz \& Soker 1994; Soker 1996;
Soker \& Zoabi 2002), in the first three classes the AGB or post-AGB stars
must be spun-up by a stellar (or substellar in the third class) companion.
In the 4th class the accretion disk is formed when the transferred mass has
high enough specific angular momentum resulting from the orbital motion.
\newline
($vi$) As stated, models for launching jets without magnetic fields exist
(e.g., Soker \& Lasota 2004).
But even if magnetic fields play a role in launching jets from accretion
disks (as most researchers think), so does the viscosity in the disk.
Still, we don't term the jet shaping viscous shaping. }}}

In young stellar objects, active galactic nuclei, and many other systems,
it is well established that accretion disks can give rise to two jets.
How exactly the two jets are launched is still an open question, but it
is also beside the point made here.
What matters is that the magnetic field might give rise to the jets,
but these are the jets which carve the PN with their action.
The geometry of the jets, whether they originate at the orbiting
companion or at the primary, their intensity and duration will
be the factors determining the resulting PN shape.
This process should be termed jet-shaping, and not magnetic-shaping.

% ==========================================================
\section{PHYSICAL REASONS WHICH ARGUE AGAINST A GLOBAL
MAGNETIC FIELD SHAPING PNS}
% ==========================================================

In this section I concentrate on three key physical characteristics of
AGB star winds and of magnetic fields which argue against global
magnetic fields playing a fundamental role in shaping PNs.

% ==========================================================
\subsection{Angular momentum considerations}
% ==========================================================

In this subsection I demonstrate that large magnetic fields will carry angular
momentum away from the stellar envelope much faster than mass is lost
by the wind. This has the effect of spinning down the star on short time
scales. Since the magnetic field is powered by rotation, this argument
disproves that such a strong large scale magnetic field is present in
the star.

At large distance from the star, any dipole field anchored to the star
(as the Earth magnetic field) is negligible. The dominant component
becomes the toroidal (azimuthal) component of the magnetic field
carried with the wind.
(The toroidal component is the magnetic field component parallel to the
equatorial plane and perpendicular both to the symmetry axis and the radial
direction).
If at some large radius the wind is compressed in the radial direction,
like in a shock wave, the toroidal component which is parallel to the
shock front, increases as the density (assuming the
magnetic field is frozen-in to the gas).
Since at large distance the toroidal component dominates, the
magnetic pressure increases as density square.
The magnetic pressure, therefore, can become larger than the thermal
pressure, as in the model of Chevalier \& Luo (1994).
In their model the wind in the transition from the AGB to the PN phase,
or the fast wind during the PN phase
{{{ (here I refer to such fields in the AGB phase as well) }}},
carries a large scale magnetic field.
This is shown schematically in figure 2 for one magnetic field line
in the equatorial plane.
Close to the star the magnetic pressure and
tension are negligible compared with the ram pressure and thermal
pressure of the wind. As the wind hits the outer PN shell, which is
the remnant of the slow wind, it goes through a shock and slows
down, and the toroidal component of the magnetic field increases
substantially. This may result in the magnetic tension and pressure
becoming the dominant forces near the equatorial plane. In particular,
the magnetic tension pulls towards the center and reduces the
effective pressure in the equatorial plane.
The force due to the tension is represented by double-line arrow
in Figure 2.
According to this model
(Chevalier \& Luo 1994; Garcia-Segura 1997), then, the equatorial
plane will be narrow, leading to an elliptical or bipolar PN.
It must be noted in regard to this and similar models, that
to globally shape circumstellar medium the toroidal (azimuthal)
field component $B_T$ must have the same sign around the symmetry axis.

Magnetic fields carry angular momentum.
{{{ Namely, the matter attached to the field lines at large distances
from the star carries angular momentum transferred to it
by the magnetic stress. }}}
This is connected to the force due to tension not having a pure radial
component, as seen in Figure 2.
Let $B_r$ be the magnetic field radial component.
The magnetic field caries an angular momentum flux (angular momentum
per unit area per unit time away from the star) of (e.g., Mestel 1968)
\begin{equation}
\dot j = \frac {\varpi B_T B_r}{4 \pi},
\label{am1}
\end{equation}
where $\varpi=r \sin \theta$ is the distance from the
rotation (symmetry) axis, and $r$ is the distance from the stellar center.
I take the magnetic field components in the equatorial plane at distance
$r$ from the stellar center from Chevalier \& Luo (1994),
\begin{equation}
B_r(r,\theta) = B_s \left( \frac{R_s}{r} \right)^2,
\label{br1}
\end{equation}
and
\begin{equation}
B_T(r,\theta) = B_s \sin \theta \frac  {v_{\rm rot}}{v_w}
\left( \frac{R_s}{r} \right)^2
\left( \frac{r}{R_s} -1  \right),
\label{bt1}
\end{equation}
where $v_w$ is the terminal wind speed, assumed to be in
the radial direction, and $\theta$ is the altitude,
measured from the polar direction ($\theta=0$, $\pi$ in the polar
directions), $B_s$ is the field at the stellar surface and $R_s$ is the
stellar radius.
One magnetic field line of this field in the equatorial plane is
drawn in Figure 2 for $v_{\rm rot}=0.1 v_w$.
On the stellar surface the field is radial.
The toroidal (azimuthal; tangential) component of the magnetic field
is created by the stellar rotation, which exerts tangential force
on the gas and magnetic field.
I consider now only the angular momentum carried by the magnetic
field; the gas itself can carry additional angular momentum from
the star.
Substituting $B_r$ and $B_T$ in equation (\ref{am1}) at
large distance from the star, $r \gg R_s$, gives
\begin{equation}
\dot j =  \frac{B_s^2 R_s \sin ^2 \theta}{4 \pi} \frac  {v_{\rm rot}}{v_w}
\left( \frac{R_s}{r} \right)^2.
\label{am2}
\end{equation}
We take the region where the magnetic field is amplified by the
stellar dynamo to be within an angle $\alpha$ from the equatorial plane.
Integrating over the outgoing magnetized wind from $\theta=\pi/2-\alpha$
to $\theta=\pi/2+\alpha$, gives the total rate
of angular momentum carried by the wind
\begin{equation}
\dot J = \int \dot j 2 \pi r^2 \sin \theta d \theta
= \left(\sin \alpha -\frac{1}{3} \sin^3 \alpha \right)
B_s^2 R_s ^3 \frac  {v_{\rm rot}}{v_w}.
\label{am3}
\end{equation}
Most of the contribution to the angular momentum
comes from regions near the equatorial plane. For $\alpha=30^\circ$
the value of the parenthesis is $\sim 0.5$, which can be considered
as the scaling value.

The field measured by the observations cited above
is the magnetic field at large distance from the star $r \gg R_s$,
where the tangential component dominate.
Let $B_{\rm obs}$ be the observed magnetic field at radius
$R_{\rm obs}$ from the star, and let us use equation (\ref{bt1})
for the tangential component, which is the component that dominates
at large radii.
We note also that the exact form we take for the variation
of the magnetic field in the wind is not important.
To globally shape the nebula the azimuthal component must
be in the same direction around the star, and this component in
a dynamo models result from the stellar rotation.
Eliminating then $B_s$ from equation (\ref{bt1}) for
$r=R_{\rm obs} \gg R_s$, and substituting in equation (\ref{am3})
for $\theta \sim 90^\circ$, gives
\begin{equation}
\dot J \simeq 0.5 B_{\rm obs}^2 R_{\rm obs}^2 R_s \frac {v_w}{v_{\rm rot}}.
\label{am4}
\end{equation}
Under the assumption of an uniformly rotating envelope, which
is likely to be the case in deep convective AGB stellar envelops,
the angular momentum of the stellar envelope is
\begin{equation}
J_s =\eta M_{\rm env} R_s v_{\rm rot},
\label{ams}
\end{equation}
where $M_{\rm env}$ is the envelope mass, and $\eta$ is defined such that
$\eta M_{\rm env} R_s^2$ is the moment of inertia of the envelope;
for AGB stars $\eta \simeq 0.2$.

As explained above, I aim to demonstrate that a star with a global
magnetic field will spin down on short time scales and this will feedback
negatively on the magnetic field which is sustained by fast rotation.
To derive a transparent expression for the spinning-down time
of the envelope, I first define the mass loss rate time scale
\begin{equation}
\tau_m  = \frac {M_{\rm env}}{\dot M},
\label{taum}
\end{equation}
where $\dot M$ is the mass loss rate from the star. This is a measure
of the time the star takes to lose its envelope.

Next, I  define the ratio of the magnetic energy density $E_B=B^2/8 \pi$ to
the kinetic energy density $E_k=\rho_w v_w^2/2$,
where $\rho_w=\dot M/4 \pi v_w R_{\rm obs}^2$ is the
wind mass density at $R_{\rm obs}$,
\begin{equation}
\chi \equiv \frac{E_B}{E_k}=
\frac{B_{\rm obs}^2 R_{\rm obs}^2}{\dot M v_w}
{{{ =3.6
\left(\frac{B_{\rm obs} R_{\rm obs}}{1 \G \AU } \right)^2
\left(\frac{\dot M}{10^{-6} \dot M_\odot \yr^{-1}} \right)^{-1}
\left(\frac{v_w}{10 \km \s^{-1}} \right)^{-1},
}}}
\label{chi}
\end{equation}
{{{ where quantities are scaled with values typical for AGB stars. }}}
Another parameter is the thermal to magnetic pressure ratio
$\beta=E_{th}/E_B$.
If either $\chi \ll 1$ or $\beta \gg 1$, magnetic field is dynamically
unimportant.
If, on the other hand $\chi$ is not too small and $\beta$ is not too large,
then the magnetic field has the potential to dominate the motion of the gas.

The spinning-down time as a result of the angular momentum
carried by the magnetic field is equal to the angular momentum of the envelope
divided by the rate at which the magnetic field carried the angular momentum
away:
\begin{equation}
\tau_j = \frac {J_s}{\dot J} \simeq
0.4 \tau_m \chi^{-1} ,
\label{tauj}
\end{equation}
where $\eta=0.2$ was substituted and where equations (6) and (7) were
substituted in and the terms rearranged to make the role of $\tau_m$ and
$\chi$ explicit.
{{{  The slowing time derived above does not refer to any
particular evolutionary phase.
However, it is most relevant to examine the phase for which models
ought to explain axisymmetrical mass loss.
This is when the AGB (or post-AGB) star loses its last $0.1-0.3 M_\odot$.
What the last equation combined with equation (\ref{taum})
show is that by the time a single star reaches this stage,
say from an envelope mass of $0.5 M_\odot$ to
and envelope mass of $0.25 M_\odot$,  it will spin extremely slowly,
below $5 \%$ of the break-up velocity.
The time scale for slowing down $\tau_j$, refers to the formation phase of
the PN shell, hence it should be compared with the mass loss time scale
$\tau_m$.
It should not be compared at all with the life time the descendant PN. }}}

{{{{ I elaborate on the meaning of equation (\ref{tauj}).
The significant quantity for the formation of the asymmetrical nebula is the fraction of
the envelope mass that is lost in a non-spherical geometry.
What equation (\ref{tauj}) shows is that after losing a small fraction of the envelope
the star will spin extremely slowly and the rest of the envelope, which contains much
more mass than the mass that has been lost in a non-symmetrical geometry,
will be lost in a spherical geometry according to the magnetic model I criticize here.
To show that, I rearrange equation (\ref{tauj}) and write it explicitly using
equations (\ref{ams}) and (\ref{taum})
\begin{equation}
\frac {d}{dt} \left(  \eta M_{\rm env} R_s v_{\rm rot} \right)
= \frac{\chi}{0.4} \frac{1}{M_{\rm env}} \frac{d M_{\rm env}}{dt}
 \left( \eta M_{\rm env} R_s v_{\rm rot} \right).
\label{tauj2}
\end{equation}
Assuming that the envelope radius and structure, as expressed in $\eta$, do not
change much during the mass loss process on the upper AGB, the last equation is simplified
to
\begin{equation}
\frac {d}{dt} \left(M_{\rm env} v_{\rm rot} \right)
\frac{dt}{d M_{\rm env}}
= \frac{\chi}{0.4}  v_{\rm rot}.
\label{tauj3}
\end{equation}
Performing the derivative gives
\begin{equation}
v_{\rm rot}+ M_{\rm env} \frac {d v_{\rm rot}}{d M_{\rm env}}
= \frac{\chi}{0.4}  v_{\rm rot},
\label{tauj4}
\end{equation}
which with the initial condition of rotational velocity $v_{\rm rot0}$ when the
envelope mass is  $M_{\rm env0}$,  has the solution
\begin{equation}
v_{\rm rot} = v_{\rm rot0}
\left( \frac{M_{\rm env}}{M_{\rm env0}} \right)^{2.5 \chi-1}.
\label{vrot}
\end{equation}
For typical values quoted in recent literature ${2.5 \chi-1} \simeq 6.5$.
Substituting this value in equation (\ref{vrot}) shows that after
only $10 \%$ of the envelope has been lost its rotation velocity drops by a
factor of $2$.
For losing $30 \%$ of the envelope the rotation velocity decreases by a factor of 10.
To summarize, what equation (\ref{vrot}) (or eq. \ref{tauj}) shows is that
independently of the mass loss time scale, the envelope spins down
substantially before most of the envelope mass has been lost.
This implies that according to the magnetic model criticized here, most of the
envelope mass will be lost in a spherical geometry.
}}}}

The slowing down time-scale of the stellar rotation will be
even shorter than that given by equation  {\ref{tauj},
because the wind mass itself carries angular momentum.
This is true for both magnetized wind closer to the equatorial plane,
and for wind flowing closer to the polar direction, where the
magnetic field was assumed to be weak.
According to equation (\ref{bt1}), as gas flows outward the
angular momentum carried by the magnetic field increases, as
the gas transfers angular momentum to the magnetic field.
When non-magnetized wind is considered, the specific angular momentum
of AGB envelopes decreases as
$l \sim M_{\rm env}^2$ (Harpaz \& Soker 1994).
What equation (\ref{tauj}) shows is that
strong large-scale magnetic fields, like those claimed in
some observations and which are required in some models,
will slow down the rotation of AGB stellar envelope on
an even shorter time scale.

Bains et al.\ (2003) find $\chi \simeq 3$ in the proto PN
OH17.7-2.0, and Bains et al.\ (2004) find $\chi \simeq 4$ in
the proto PN IRAS 20406+2953.
Vlemmings et al.\ (2002) find the thermal to magnetic
pressure to be $\beta=0.05$, for a gas
at a temperature of $1000 \K$ in H$_2$O maser clumps;
from that and for a wind velocity of $10 \km \s^{-1}$
I find $\chi=5$.
In the H$_2$O maser clumps in the wind of the red
supergiant S Per, Richards et al.\ (2005) find the
thermal to magnetic pressure to be $\beta=0.05$, for a gas
at a temperature of $1000 \K$. For a wind velocity of
$10 \km \s^{-1}$ I find $\chi=5$.
In any case, in S Per the magnitude and direction
of the magnetic field measured from OH masers contains
no large scale structure, hence, cannot serve for
large scale shaping.
It is more likely consistent with local magnetic clouds,
as drawn in figure 1 (Soker \& Kastner 2003).
Bains et al.\ (2003, 2004) and other authors conclude that $\chi$$>$1
is evidence for global shaping of the circumstellar gas by magnetic fields.
However, when seen in the context of equation {\ref{tauj}, $\chi> 1$ simply
points to the fact that the spin-down time $\tau_j$ is smaller than the
time-scale over which the entire envelope is ejected.
In other words, if such a field exists, the star will dramatically slow down
its rotation and quench its global field before the field has time to shape the
mass-loss.
Without a continuous supply of angular momentum, the
star will not be able to sustain a strong tangential field
which is required to shape the wind in magnetic shaping models.
Such a source of angular momentum, could easily be in the form of
a binary companion (the small giant's core cannot store enough
angular momentum).
But once we invoke the presence of a companion other effects will come
into play which will outperform the action of the magnetic field
e.g., by accreting mass and blowing jets (Soker 2004).
{{{ I emphasize that the need for a companion comes from the short
spinning-down time, namely, the star substantially slows down before
much of the envelope is lost $\tau_j \ll \tau_m$.
The spin-down results from the angular momentum carried by the gas
in the wind (e.g., Harpaz \& Soker 1994), or from the angular
momentum due to the magnetic stress as shown in the present paper.
The inequality $\tau_j \ll \tau_m$ which implies the need for a companion,
is better fulfilled in dynamical magnetic models when $\chi \ga 1$,
as evident from equation (\ref{tauj}).
Without a companion, then, the star will spin much too slow to
expel axisymmetrical wind; a spherical rather than bipolar or elliptical
PN will be formed without a companion.  }}}

In another type of model (e.g., Matt et al.\ 2000 for PNs;
Matt \& Balick 2004 for the extreme massive star $\eta$ Carinae)
the star has a very strong dipole magnetic field.
In the model of Matt \& Balick (2004) the magnetic field pressure on the
photosphere is {{{ globally (not locally as in solar spots) }}} larger than the
thermal pressure (which makes their value of the magnetic field suspiciously
unrealistic).
The very strong magnetic field keeps the wind in corotation with the
envelope to several stellar radii.
Hence, the specific angular momentum of the wind is very large,
much more than that of a wind in non-magnetized stars.
So, even if such large, frozen in, dipole magnetic fields did exist,
from angular momentum considerations it is clear that the star slows
down on a very short time scale, such that the
magnetic field could not be supported anymore.

% ==========================================================
\subsection{Energy considerations}
% ==========================================================
Another way to look at the requirement for a stellar companion
in models based on dynamically important magnetic fields
is from energy considerations.
Away from the star the dominant field has to be tangential as argued
in the last subsection; hence the observed milli-Gauss fields, which are
known to originate far from the stellar surface cannot be large scale
{\it radial} fields.
As we have discussed, the tangential field is created by rotation.
The energy carried in the tangential component of the field within
a wind mass element $\Delta M_w$ is
\begin{equation}
\Delta E_B= \frac{\chi}{2} \Delta M_w v_w^2,
\label{eb1}
\end{equation}
where as before, $\chi$ is the ratio of magnetic to kinetic energy density.
The total rotational energy of the envelope is
\begin{equation}
E_{\rm env} = \frac{\eta}{2} M_{\rm env} v_{\rm rot}^2,
\label{ebe}
\end{equation}
where a uniform envelope rotation is assumed, and, as before,
$v_{\rm rot}$ is the equatorial rotation velocity.
The fraction of the envelope mass that can be lost within a
strong magnetic field is given by equating the energies in the
last two expression
\begin{equation}
\frac {\Delta M_w}{M_{\rm env}} =  \frac{\eta}{\chi}
\left( \frac {v_{\rm rot}}{v_w} \right)^2.
\label{dm1}
\end{equation}
For AGB stars $\eta \sim 0.2$.
Therefore, in order to expel at least $\sim 10\%$ of the AGB envelope
mass with a strong magnetic field of $\chi \sim 1-3$, the
envelope equatorial rotation velocity must be
$v_{\rm rot} \ga v_w \simeq 10 \km \s^{-1}$.
Such a rotation velocity is not indigenous to {{{ singly evolved }}}
AGB stars and requires the AGB to be spun-up by a companion, at
least as massive as a brown dwarf (Harpaz \& Soker 1994).

 In the numerical calculations of  Garc\'{\i}a-Segura et al.\
(2005) an AGB star of radius $4.5 \AU$ expels toroidal magnetic field,
with total ejected energy of $\sim 10^{46} \erg$ after $\sim 1000 \yr$.
As the global toroidal field results from rotation, the
parameters of such an envelope are impossible to meet
for an AGB star; using equation (\ref{ebe}), an envelope
mass of $1 M_\odot$ is required to rotate with an equatorial
velocity of $70 \km \s^{-1}$$-$above the break-up speed.
Even for a binary companion to spin-up the envelope such parameters
are on the extreme.

% ==========================================================
\subsection{Deceleration by magnetic stress}
% ==========================================================
In some magnetic field shaping models (model family (2) in section
2.1), equatorial magnetic stress slows down the wind expansion in the
equatorial plane, leading to the formation of an elliptical PN
(Chevalier \& Luo 1994).
In the Chevalier \& Luo (1994) model, the stress becomes
important at late times, when the post-AGB wind, which carries the
magentic field,  hits the previously ejected AGB wind.
This is not the case for the claimed value of the magnetic fields cited
in section 2.1, where the magnetic to kinetic energy density being
$\chi \ga 1$ already in the AGB or proto-PN phase.
For a global shaping, the tangential component of the magnetic field,
which is the dominant component, must encircle the star.
The stress force per unit volume in the equatorial plane,
pulling toward the center,  is
\begin{equation}
f =  \frac{B^2}{4 \pi R}.
\label{f1}
\end{equation}
The deceleration of the wind (or acceleration toward the center) is
\begin{equation}
a=\frac{f}{\rho}=\chi \frac {v_w^2}{R}.
\label{a1}
\end{equation}
The time required to stop the wind's expansion is
\begin{equation}
\tau_w=\frac{v_w}{a} = \frac{R}{v_w} \chi^{-1}
{{{ = 160 \left(\frac{R}{1000 \AU} \right)
\left(\frac{v_w}{10 \km \s^{-1}} \right)^{-1}
\left(\frac{\chi}{3 \AU} \right)^{-1}  \yr. }}}
\label{a2}
\end{equation}
For the claimed value of $\chi \simeq 3$ made by, e.g., Bains et al.\
(2003, 2004) (see subsection 3.1), the wind should have stopped long before it
reaches its present observed maser spots, which requires a time $R/v_w$.
In any case, this region should not expand, but rather move
toward the center.
This is in contradiction with the observed expansion velocity
of $\sim 14 \km \s^{-1}$ found by Bains et al.\ (2003)
in the proto-PN OH17.7-2.0, where they also find $\chi \sim 3$.

At large distances from the star the tangential component
becomes the dominant one, as is assumed throughout the paper.
It is not possible that the observed magnetic field at
$r \sim 1000 \AU$ and of magnitude $\sim 10^{-3} \G$ is a large scale
radial field.
This is because the radial component decreases as $r^{-2}$, which implies
that the magnetic field on the photosphere at $R_s \sim 1 \AU$ is
$B_s \sim 1000 \G$ and the magnetic pressure
$P_{Bs} \sim 4 \times 10^4 \erg \cm^{-3} \ga 10 P_{ths}$, where
$P_{ths}$ is the photospheric thermal pressure, which for AGB
stars is $\sim 10^3 \erg \cm^{-3}$.
For small magnetic loops, on the other hand, local evolution,
such as magnetic stress, can make the radial component comparable
to the tangential one.

% ==========================================================
\subsection{Ambipolar diffusion}
% ==========================================================

The magnetic field lines carry the ionized particles, and
diffuse through the neutral particles in a process termed
ambipolar diffusion. If ambipolar diffusion is taking place
the effect of the magnetic field in directing the motion of the
gas and determining the PN shape is substantially lessened.
In this section we show that AGB winds are within the ambipolar
diffusion regime and that this constitutes another reason,
although not the main one, why global
field shaping does not work.

Bains et al. (2004) use the ambipolar diffusion time as given by
Hartquist \& Williams (1989). However, Hartquist \& Williams (1989)
study self gravity clouds, supported by the magnetic field, i.e.,
the force due to magnetic pressure equals gravity.
This is far from the situation in AGB stars, where
away from the stellar surface gravity is negligible.
Therefore, Bains et al. (2004) derive an ambipolar time scale
several orders of magnitude too long.

The ambipolar drift velocity between ions and neutral particles
is given by $v_D=B^2(4 \pi \Gamma \rho_i \rho_n R)^{-1}$,
where $\rho_i$ and $\rho_n$ are the densities of the ions and
neutral particles, respectively, $R$ is the curvature radius
of the magnetic field, while $\Gamma$ is a coefficient which
depends on the drift velocity itself (e.g., Shu et al.\ 1987).
At the low temperatures dominating the maser clumps, the ions frozen
to the field are mostly from singly ionized metals, rather than hydrogen
or helium, which have a much higher ionization potential.
For the conditions relevant to the maser clumps, the
drift velocity is
\begin{equation}
v_d \simeq 1
\left(\frac {B}{10^{-3} {\rm G}} \right)^2
\left(\frac {R}{10^{16} \cm} \right)^{-1}
\left(\frac {n_i n_n}{10^4 \cm^{-2}} \right)^{-1}
\left(\frac {\Gamma}{10^{14} \cm^3 \g^{-1} \s^{-1}} \right)^{-1}
\km \s^{-1},
\label{amp1}
\end{equation}
where the ion ($n_i \sim 1 \cm^{-3}$) and neutral particles
($n_n\sim 10^4 \cm^{-1}$) number densities were
scaled according to Bains et al.\ (2004).
Equation \ref{amp1} demonstrates that the ambipolar diffusion speed can be
comparable to the wind speed for a magnetic field of $B \sim 3 {\rm mG}$.
The {{{ wind }}} density decreases as $r^{-2}$,  {{{ and it is assumed that
the same dependance holds in maser spots, hence $n_i$ and $n_n$
decreases as $r^{-2}$. }}}
The tangential magnetic component, on the other hand,
decreases as $r^{-1}$.
Therefore, the ambipolar speed increases as $\sim r$, and cannot
be neglected in models where magnetic fields are strong.
In the ambipolar diffusion process the neutral gas leaves the
magnetic fields and is not influenced by it, hence PN shaping cannot
result in the ambipolar regime.
I should also notice, in passing, that the acceleration of the magnetic field
lines toward the center (previous section) increases.

In the local magnetic field geometry (model 3 in section 2.2) we find
different parameters.
Magnetic fields are expected to be weaker, and their radius of
curvature smaller. The ambipolar diffusion might be important in this
regime (Equation~\ref{amp1}), and the magnetic field
has enough time to further arrange itself, via stress and diffusion,
in local clumps. These clumps are observed as dense maser clumps.
As discussed by Soker (2002), the magnetic pressure of
small scale magnetic fields (e.g., loops) can be enhanced
at large distances from the star, and become larger than the
thermal pressure.
But it is in small isolated regions that the magnetic fields are strong,
hence they cannot globally shape the nebula.

% ==========================================================
\section{DISCUSSION AND SUMMARY}
% ==========================================================

In the course for the present work, I examined recent claims for strong
magnetic fields around evolved stars (e.g.,
Miranda et al.\ 2001; Vlemmings et al.\ 2002, 2005; Diamond \& Kemball 2003;
Bains et al.\ 2003, 2004a,b; Szymczak \& Gerard 2004, 2005;
Jordan et al.\ 2005), and the conclusion by some authors that these
magnetic fields have had a role in shaping their PNs.
By building on the arguments of Soker \& Zoabi (2002) and Soker (2004) I
expose additional reasons why these conclusions are unjustified.

In order to influence the large scale structure of the circumstellar
matter (the wind), the magnetic field must have a large scale structure.
However, I showed (section 3.1) that if the observed magnetic fields have
a large scale structure, then they contain more angular momentum and energy
than a single star can supply.

I have also considered deceleration by magnetic stress.
Here the toroidal field around the AGB star would contribute
to decelerating the wind at the equator, possibly reversing its motion.
This is contrary to the observation of the wind expansion velocity.
Finally, I have determined the ambipolar diffusion speed for the regimes
likely to operate in AGB stars.
AGB stellar winds, if contain magnetic field of milli-Gauss at
$r \sim 1000 \AU$, are well within the ambipolar diffusion regimes
and the drifting of the field through the gas is significant.
This diminishes the power of the magnetic field action on the
gas because it effectively decouples the gas from the field.

I also discussed that any claim that the strong observed magnetic
fields demonstrate that the PN was shaped by them, must consider
several physical characteristics of the system that can make
global fields implausible PN shaping agents.
For example, as mentioned here in some cases the surface magnetic
pressure is larger than the thermal pressure on the entire surface
(and not locally as in the Sun); this does not make
sense as the photosphere of the star would be unstable.
As discussed in sections 2 and 3, the models for large
scale magnetic shaping must consider the stress of the magnetic field.
When this is done, some contradictions between observation
and the expected behavior of the field appear.

By demonstrating that the magnetic fields observed cannot play a global
role in shaping the PN, I am not denying that magnetic fields can exist
in AGB stars and their progeny. These observed fields
can be attributed to local ejection events, similar to the magnetic clouds
in the solar wind (Soker \& Kastner 2003; fig. 1 here).
Such locally-enhanced magnetic fields can be formed by a dynamo mechanism
rooted mainly in the vigorous convective envelope of AGB stars.
For this mechanism to operate the AGB star does not need to rotate fast;
slow rotation is sufficient and can be achieved even by an orbiting
planet spinning up the envelope.
Local magnetic fields can influence the wind geometry by
facilitating dust formation.
Locally enhanced magnetic fields on the AGB stellar surface can
lead to the formation of cool
spots and cool filaments, as is the case in the Sun.
Dust formation rate, hence mass loss rate, is enhanced above these cool
spots and within the cool filaments (Soker \& Zoabi 2002).
If spots are concentrated near the equator, then mass loss rate is
higher in that direction.
This process might lead to the formation of moderately elliptical PNs
(those with small departure from sphericity), but
cannot account for lobes, jets, etc. Hence, these local magnetic fields
can play a role in the PN morphology but this is a secondary role
to that exercised by companions to the AGB star.

Ignoring the effects of binary companions in the shaping of PNs and related
nebulae, will lead to erroneous conclusions. For instance,
Jordan et al.\ (2005) argue that their discovery of magnetic
field on the surface of four central stars supports the
hypothesis of magnetic shaping.
In the case of systems like EGB 5 (PN G211.9+22.6), one of the four systems
they discuss, the known WD companion (Karl et al. 2003)
is expected to influence the mass loss geometry more than any
possible magnetic field in AGB stars.

As I have stated and demonstrated in previous papers
(Soker 2004, and references there in) the effect of companions
to the AGB stars, even as small as brown dwarfs and planets,
can give rise to a host of physical phenomena, resulting in shaping
of the AGB mass-loss and subsequent wind into the morphologies observed.

\acknowledgments
I thank Orsola De Marco for many useful comments and for
substantially improving the original manuscript,
and Vikram Dwarkadas for useful suggestions.
This research was supported in part by the
Israel Science Foundation.

% ============ 1 FFFFFFFFFFFFFFFFFFFFFFFFFFFFFFFFFFFFFFFFFFFFF
% ============ 1 FFFFFFFFFFFFFFFFFFFFFFFFFFFFFFFFFFFFFFFFFFFFF
%\epsscale{.10}
\begin{figure}
\includegraphics[width =105mm]{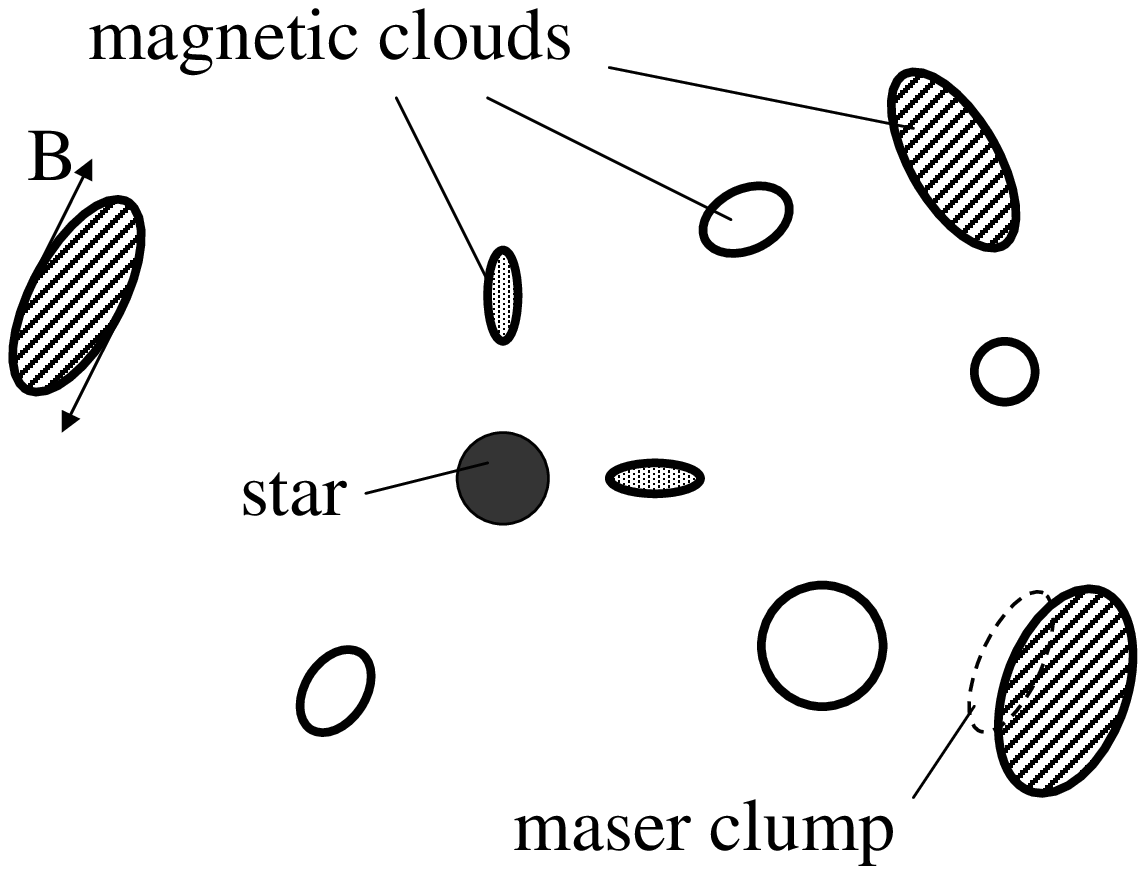}
% \vskip 0.2 cm
%%\includegraphics[width =135mm]{nomagnetff11.pdf} \vskip 0.2 cm
% made from nomagnetff1a.doc
\caption{Schematic drawing of the magnetic field topology as
suggested in the paper. The magnetic flux loops and the gas frozen
to them are termed magnetic clouds. Closer to the star the radial
component dominate, and the loops are radially elongated
(represented by the two gray loops). At larger distances the
azimuthal component dominate, and the loops may become
tangentially elongated (represented by the three loops with
diagonal lines). Magnetic stress tends to make the loops more
circle. Magnetic flux loops cannot shape PNs on large scale.
Magnetized maser spots might cover only a fraction of a loop,
where conditions are favorable for masing operation, hence will
show coherence magnetic field on a small scale (shown schematically
by the dashed circle)}
\end{figure}
% ============ 1 FFFFFFFFFFFFFFFFFFFFFFFFFFFFFFFFFFFFFFFFFFFF
% ============ 1 FFFFFFFFFFFFFFFFFFFFFFFFFFFFFFFFFFFFFFFFFFFFF
% Made from jetfig1w.doc
%\epsscale{.20}
\begin{figure}
\includegraphics[width =105mm]{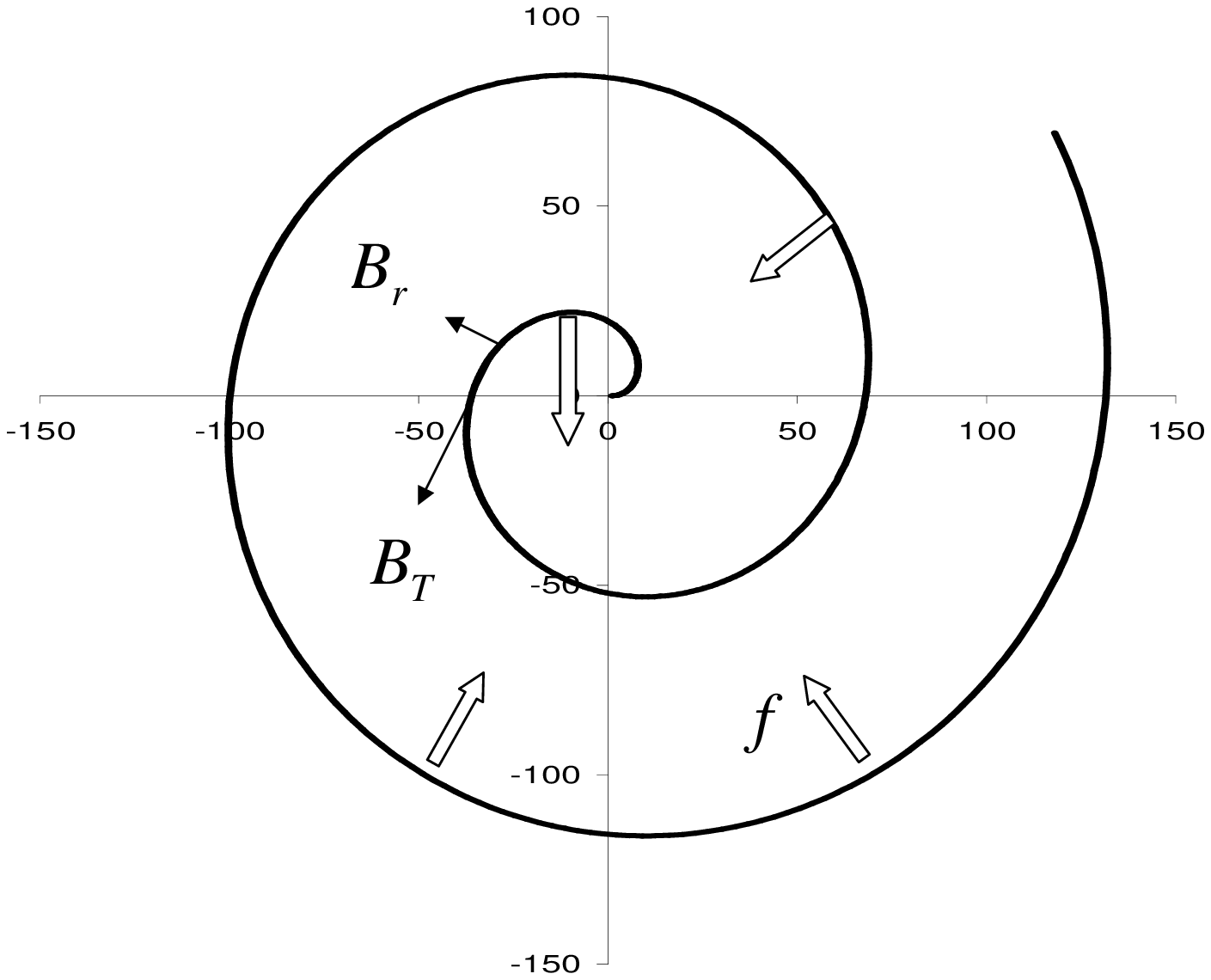} \vskip 0.2 cm
% made from nomagnetff1.doc
\caption{The structure of a filed line in the equatorial
plane ($\theta=\pi/2$) according to equations {\ref{br1}}
and {\ref{bt1}}, for wind velocity ten times the
equatorial rotation velocity, $v_{\rm rot}=0.1v_w$.
The double-line arrows represent the the force on the
gas due to the magnetic stress.
this force is not purely radial.
The axes are marked in units of the stellar radius ($R_s$).}

\end{figure}
% ============ 1 FFFFFFFFFFFFFFFFFFFFFFFFFFFFFFFFFFFFFFFFFFFF
\end{document}